\documentclass[12pt]{iopart}

\newcommand{\AuAu }{Au+Au }
\newcommand{\dAu }{$d$+Au }
\usepackage{iopams}
\usepackage{graphicx}
\begin{document}

\title[Away-side Modification and Near-side Ridge Relative to Reaction Plane] {Reaction Plane Dependent Away-side Modification
and Near-side Ridge in \AuAu Collisions}

\author{Aoqi Feng (for the STAR collaboration)}

\address{Institute of Particle Physics, Huazhong Normal University, Wuhan, China, 430079
Department of Physics, Purdue University, West Lafayette, Indiana,
USA, 47907} \ead{afeng@purdue.edu}

\begin{abstract}
STAR preliminary results of di-hadron correlations versus
$\phi_{s}$, the trigger particle azimuthal angle relative to the
constructed event plane are reported for mid-central \AuAu
collisions and compared to central \AuAu as well as minimum bias
\dAu collisions. The correlations are observed to vary with
$\phi_{s}$ on both the near and away side of the trigger particle.
The away-side correlation evolves from single- to double-peak with
increasing $\phi_{s}$. The near-side correlation is separated into
`jet' and `ridge': the ridge is found to decrease with $\phi_{s}$
while the jet remains relatively constant.
\end{abstract}


\section{Introduction}
Di-hadron correlations provide a valuable probe to study the
medium created at RHIC. Their measurements have revealed strong
suppression of back-to-back azimuthal correlation at high
transverse momentum ($p_{T}$) consistent with
jet-quenching~\cite{backToback}, enhancement and strong
modification at low $p_{T}$ suggesting strong jet-medium
interactions~\cite{Fuqiang}, and the formation of a long range
correlation in pseudorapidity------the so-called
ridge~\cite{Fuqiang,Jorn}.

Moreover, high $p_T$ di-hadron correlation study relative to the
constructed event plane (EP) has shown a stronger back-to-back
suppression when the trigger particle is out-of-plane than
in-plane~\cite{Inout}. In this work, we extend this study to finer
slices in $\phi_{s}$, the trigger particle azimuthal angle
relative to EP and to lower associated $p_{T}^{assoc}$. The
20-60\% mid-central \AuAu collisions at 200 GeV are used, and are
compared to the top 5\% central \AuAu as well as minimum bias \dAu
collisions. The $p_{T}$ ranges of the trigger and associated
particles are $0.15<p_{T}^{assoc}<3<p_{T}^{trig}<4$ GeV/$c$, and
their pseudorapidity range is restricted to $|\eta|<1$.

\section{Analysis and Systematic Uncertainties}
The di-hadron correlation structure sits atop a large flow
background given by~\cite{background}

\begin{equation}
\frac{dN_{bgkd}^{pairs}} {d\Delta\phi} = B\left[
1+2v_{2}v_{2}^{R}\cos(2\Delta\phi)+2v_{4}v_{4}^{R}\cos(4\Delta\phi)\right]
\label{Equation:v4Background}
\end{equation} where $v_{2}$ and $v_{4}$ are the associated particle anisotropic
flow, and $v_{2}^{R}$ and $v_{4}^{R}$ are the average anisotropy
of trigger particles within slice
$\phi_{s}-c<|\phi_{trig}-\Psi_{EP}|<\phi_{s}+c$. The event plane
angle $\Psi_{EP}$ is constructed from particles outside the
respective $p_T$ range used in each correlation study. We consider
the flow background up to the order of $v_{2}v_{4}$ which is about
10\% of the final correlation signal. Namely,
\begin{eqnarray}
\label{Equation:v2Rv4R} v_{2}^{R} =
\frac{T_{2}+(1+T_{4})v_{2}^{trig}+(T_{2}+T_{6})v_{4}^{trig}}{1+2T_{2}v_{2}^{trig}+2T_{4}v_{4}^{trig}},\quad
v_{4}^{R}=\frac{T_{4}+(T_{2}+T_{6})v_{2}^{trig}}{1+2T_{2}v_{2}^{trig}},
\end{eqnarray} where $T_k =
\cos(k\phi_s)\frac{\sin(kc)}{kc}\langle\cos(k\Delta\Psi)\rangle$
($k$=2,4), and $v_2^{trig}$ and $v_4^{trig}$ are trigger particle
flow.

The systematic uncertainties on our results come mainly from
background subtraction which are dominated by uncertainties in the
flow measurements and background normalization by $B$ in
Eq.(\ref{Equation:v4Background}). We use the average of $v_{2}$
from the modified reaction plane ($v_2\{MRP\}$) and the 4-particle
cumulant ($v_2\{4\}$) method and use the range between the two as
our systematic uncertainty~\cite{Fuqiang}. We use the
parameterized $v_{4}=1.15v_{2}^{2}$ for $v_4$ values and the error
calculation. We assume zero signal in the lowest $\Delta\phi$
region of size 0.52 to obtain $B$. We vary the $\Delta\phi$ size
between 0.26 and 0.78 to estimate the systematic uncertainty on
$B$.

\section{Results and Discussions}
\begin{figure}[th]
\centering\includegraphics[width=0.9\textwidth,height=0.6\textwidth]{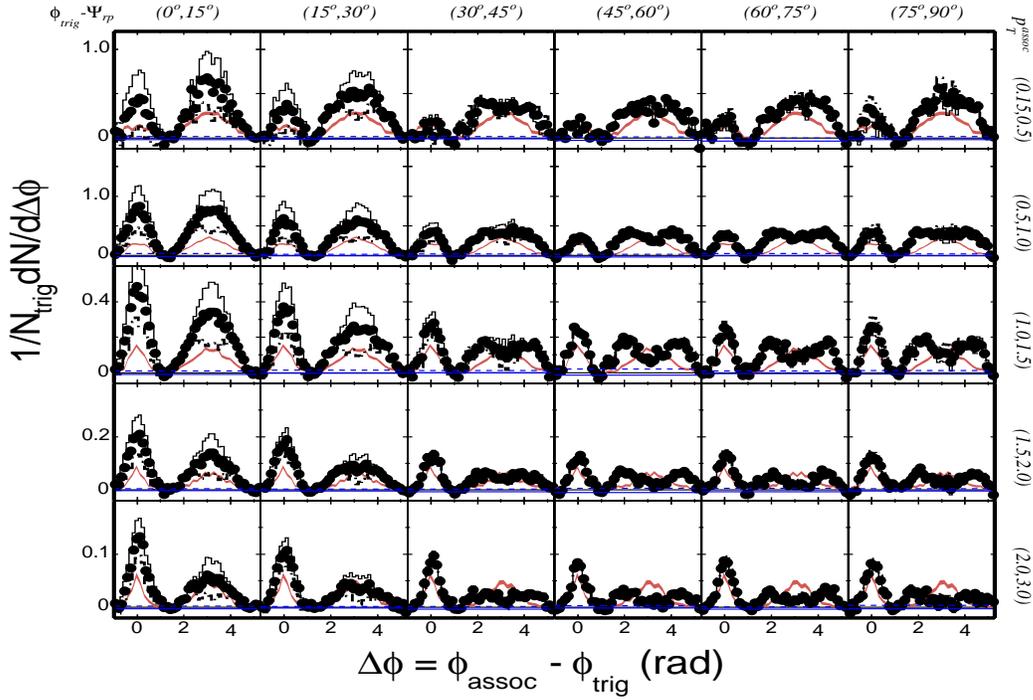}
\caption{Background subtracted di-hadron correlations as a
function of $\phi_{s}=\phi_{trig}-\Psi_{EP}$ and $p_{T}^{assoc}$
for $3<p_{T}^{trig}<4$ GeV/$c$ in 20-60\% \AuAu collisions. The
$\phi_s$ range increases from $0$-$15^{\circ}$ (left column) to
$75$-$90^{\circ}$ (right column); the $p_{T}^{assoc}$ range
increases from 0.15-0.5 GeV/$c$ (top row) to 2-3 GeV/$c$ (bottom
row). The histograms and dashed lines indicate the systematic
uncertainties from flow and background normalization,
respectively. The mini-bias \dAu inclusive di-hadron correlation
is superimposed in red for comparison.} \label{6slice_5x6T3}
\end{figure}

\begin{figure}[th]
\begin{minipage}[c]{0.5\textwidth}\centering\mbox{
\includegraphics[width=0.8\textwidth]{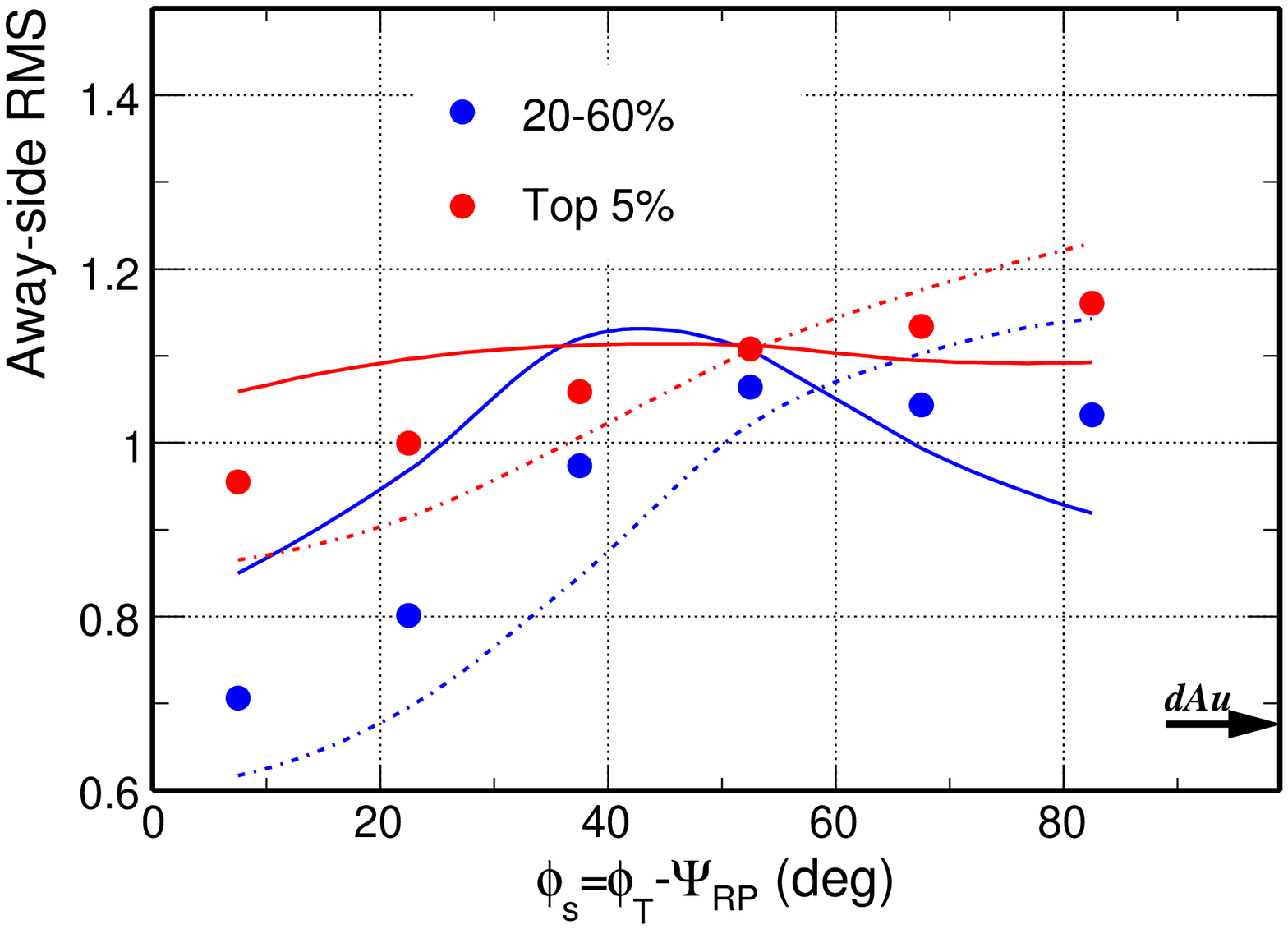}}
\end{minipage}
\begin{minipage}[c]{0.5\textwidth}
\includegraphics[width=0.8\textwidth]{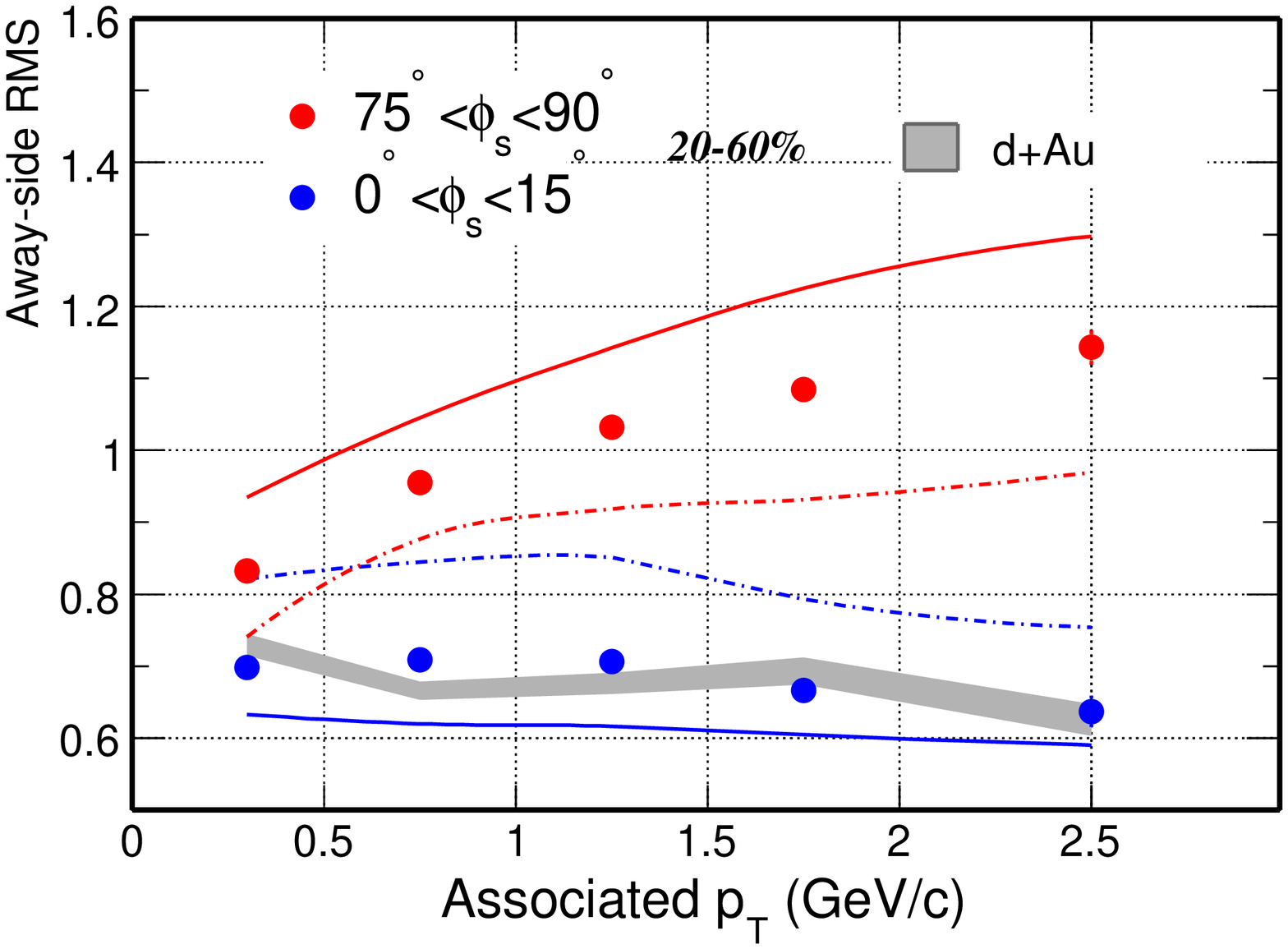}
\end{minipage}
\caption{Left panel: The di-hadron correlation function away-side
RMS versus $\phi_{s}$ in 20-60\% (blue) and top 5\% (red) \AuAu
collisions for $1.0 <p_{T}^{assoc}< 1.5$ GeV/$c$. Right panel: The
away-side RMS for two $\phi_s$ slices versus $p_{T}^{assoc}$ in
20-60\% \AuAu collisions. The trigger $p_T$ range is $3 <
p_{T}^{trig} < 4$ GeV/$c$ for both panels. The curves indicate
systematic uncertainties due to flow (solid: $v_{2}\{4\}$; dashed:
$v_{2}\{MRP\}$).The corresponding \dAu result is indicated by the
arrow and the shaded area.} \label{RMS}
\end{figure}

Figure \ref{6slice_5x6T3} shows the background subtracted
di-hadron correlation as a function of $\phi_{s}$ and
$p_{T}^{assoc}$. The away-side correlation evolves dramatically
from single-peak to double-peak as $\phi_{s}$ increases from
in-plane ($\phi_{s}\sim 0^{\circ}$) to out-of-plane ($\phi_{s}\sim
90^{\circ}$) for most $p_T^{assoc}$ bins. To quantify the
away-side modification, we show in Fig. \ref{RMS} (left) the RMS
of the correlation function within $|\Delta\phi-\pi|<\pi-1$ versus
$\phi_{s}$ in 20-60\% as well as the top 5\% \AuAu data. The RMS
increases with $\phi_{s}$ (i.e. the away-side distribution
broadens with $\phi_{s}$); the effect is smaller in central
collisions. For $\phi_{s}\sim 0^{\circ}$ the RMS in 20-60\%
collisions is not much larger than in \dAu, while in 5\% central
collisions, it already shows a marked broadening from $d$+Au. This
is qualitatively consistent with the different pathlengths the
away-side parton traverses in the reaction plane (RP) direction
for the two centralities. On the other hand, the RMS for
$\phi_{s}\sim 90^{\circ}$ are not much different between the two,
again consistent with the collision geometry--- the pathlengths
perpendicular to RP are similar. The $p_{T}^{assoc}$ dependence of
the away-side RMS is shown in Fig. \ref{RMS} (right) for the
20-60\% centrality. The RMS remains constant over all
$p_{T}^{assoc}$ at small $\phi_{s}$, and increases with
$p_{T}^{assoc}$ at large $\phi_{s}$. The double-peak structure is
stronger when the trigger particle is further away from RP and the
associated particle is harder.

The shape of the near-side correlation shown in Fig.
\ref{6slice_5x6T3} remains relatively unchanged, while the
amplitude drops with $\phi_{s}$ and then appears to saturate for
all $p_{T}^{assoc}$. Motivated by previous
observations~\cite{Fuqiang,Jorn}, we separate the near-side
correlation into `jet' and `ridge' by analyzing the data in two
$\Delta\eta$ regions: $|\Delta\eta|>0.7$ where the ridge is the
dominant contributor and $|\Delta\eta|<0.7$ where both jet and
ridge contribute. The jet part is in turn obtained from the
difference between the raw correlation in $|\Delta\eta|<0.7$ and
that in $|\Delta\eta|>0.7$ scaled by the $\Delta\eta$ acceptance
factor of approximately 1.45. The systematic uncertainties due to
background and flow are largely cancelled in the jet result
because flow is measured to be independent of $\eta$ in our
measured range.

\begin{figure}[th]
\begin{minipage}[c]{0.5\textwidth}
\includegraphics[width=0.8\textwidth]{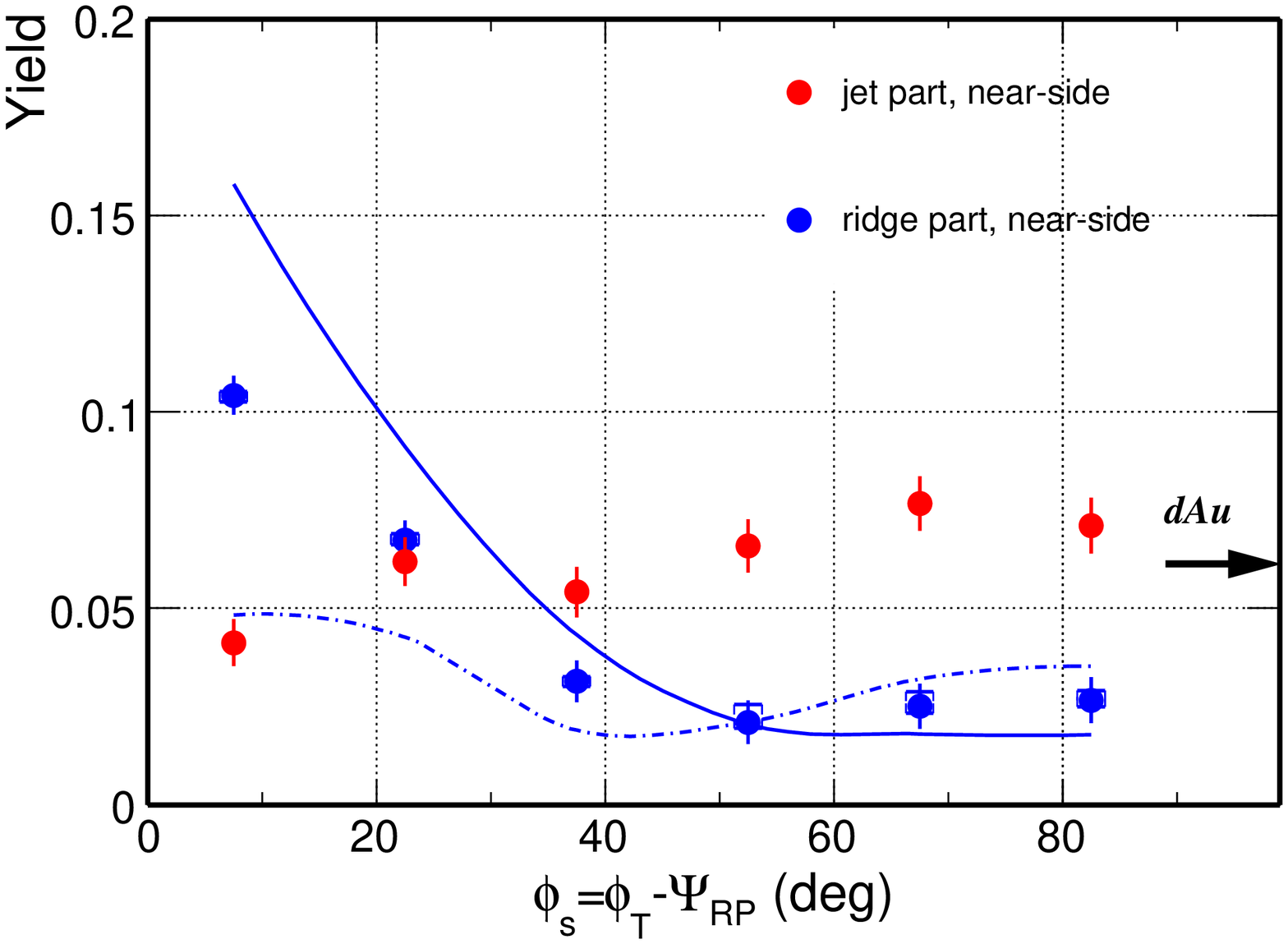}
\end{minipage}
\begin{minipage}[c]{0.5\textwidth}
\includegraphics[width=0.8\textwidth]{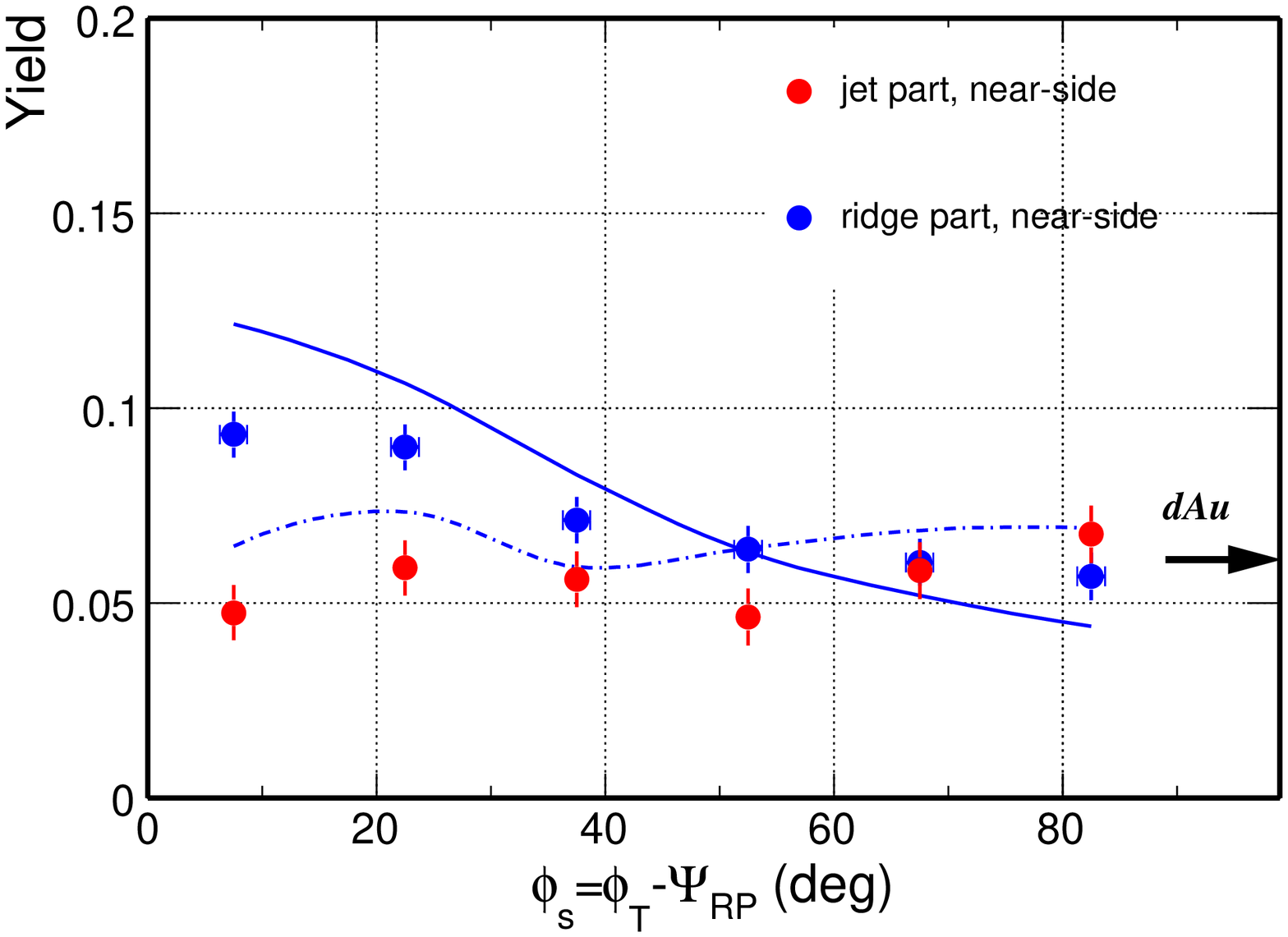}
\end{minipage}
\caption{The near-side di-hadron correlation yields in
$|\Delta\eta|<0.7$ and $|\Delta\phi|<1$ for the jet part (red) and
the ridge part (blue) as a function of $\phi_{s}$. The results are
for $3<p_{T}^{trig}<4$ GeV/$c$ and $1.5<p_{T}^{assoc}<2.0$ GeV/$c$
in 20-60\% (left) and top 5\% (right) \AuAu collisions. The curves
indicate systematic uncertainties (solid: $v_{2}\{4\}$, dashed:
$v_{2}\{RP\}$). The d+Au result is indicated by the arrows.}
\label{JetRidge}
\end{figure}

Figure \ref{JetRidge} shows the jet and ridge yields in
$|\Delta\eta|<0.7$ and $|\Delta\phi|<1$ as a function of
$\phi_{s}$ for both 20-60\% and top 5\% collisions. While the jet
yield remains relatively constant (or slightly increases), the
ridge yield decreases with $\phi_{s}$, more significantly in the
20-60\% centrality. The data indicate that the ridge is
predominant only in the RP direction in mid-central collisions,
presumably due to strong interactions between the near-side parton
and the medium. The near-side jet perpendicular to RP, on the
other hand, suffers minimal interaction, resulting in no
significant ridge. The $\phi_s$ dependence of the ridge yield in
the top 5\% collisions is weaker, qualitatively consistent with
the more spherical collision geometry in these collisions. It is
also interesting to note that the ridge yields at small $\phi_{s}$
are similar between the two centralities, perhaps due to the
similar surface curvature and/or gluon density in the RP direction
for the two centralities.

\section{Summary}

We have reported the STAR preliminary results of di-hadron
correlations as a function of $\phi_s$, the trigger particle
azimuthal angle relative to the constructed event plane in \AuAu
collisions at 200 GeV. The correlations depend on $\phi_s$ on both
the near and away side. The away-side structure in 20-60\%
collisions evolves from single- to double-peak with trigger
particles from in-plane to out-of-plane, while the evolution is
less significant in central collisions where the double-peak
structure is evident already at small $\phi_{s}$. At large
$\phi_s$ no significant difference on the away side is observed
between the two centralities. The away-side data suggest
pathlength effect in jet-quenching.

The near side is decomposed into jet and ridge. The jet yield
remains relatively constant over (or slightly increases with)
$\phi_s$. The ridge yield is significant at small $\phi_s$ and
drops with increasing $\phi_s$ in 20-60\% \AuAu collisions. The
ridge yields in the reaction plane direction are similar for the
two centralities. In the direction perpendicular to the reaction
plane, sizeable ridge remains in central collisions, while little
ridge is observed in 20-60\% collisions. These results may suggest
connection of the surface geometry and the formation to the
near-side ridge.

\end{document}